\documentclass[twocolumn]{webofc}
\usepackage[varg]{txfonts}   
\usepackage{units}   
\usepackage{lineno}
\begin{document}
\title{Search for a correlation between the UHECRs measured by the Pierre Auger Observatory and the Telescope Array and the neutrino candidate events from IceCube and ANTARES}
%
%

\author{\firstname{J.} \lastname{Aublin, }\inst{1}\firstname{ A.} \lastname{Coleiro, }\inst{1,2}  \firstname{ A.} \lastname{Kouchner }\inst{1,3} \; for the ANTARES Collaboration 
\newline \firstname{I.} \lastname{ Al Samarai,}\inst{4} \firstname{ A.} \lastname{Barbano, }\inst{4}  \firstname{ T.} \lastname{Montaruli }\inst{4} \firstname{ L.} \lastname{Schumacher}\inst{5} \firstname{ C.} \lastname{Wiebusch
}\inst{5} \; for the IceCube Collaboration
\newline \underline {\firstname{L.} \lastname{Caccianiga,}\inst{6}} \thanks{\email{lorenzo.caccianiga(at)unimi.it}} \firstname{ P.L.} \lastname{Ghia, }\inst{7}\firstname{ U.} \lastname{ Giaccari,}\inst{8} \firstname{ G.} \lastname{Golup }\inst{9}\;for the Pierre Auger Collaboration
\newline \firstname{H. } \lastname{Sagawa,}\inst{10} \firstname{ P.} \lastname{ Tinyakov}\inst{11}\; for the Telescope Array Collaboration}

\institute{APC, Univ Paris Diderot, CNRS/IN2P3, CEA/Irfu, Obs de Paris, Sorbonne Paris Cité, France \and 
IFIC - Instituto de Fìsica Corpuscular (CSIC - Universitat de València) c/ Catedràtico José
Beltràn, 2 E-46980 Paterna, Valencia, Spain \and Institut Universitaire de France, 75005 Paris, France \and Départment de physique nucléaire et corpusculaire, Université de Genève, Genève, Switzerland. \and RWTH Aachen University, III. Physikalisches Institut B, Aachen, Germany. \and Dipartimento di Fisica "Aldo Pontremoli", Università degli Studi di Milano e INFN, Milan, Italy \and Institut de Physique Nucl\'eaire d'Orsay (IPNO), CNRS-IN2P3, Orsay, France \and Centro Brasileiro de Pesquisas Fìsicas, Rio de Janeiro, RJ, Brazil  \and Centro Atómico Bariloche, S.C de Bariloche, Argentina \and Institute for Cosmic Ray Research, University of Tokyo, Kashiwa, Chiba, Japan \and Service de Physique Théorique, Université Libre de Bruxelles, Ixelles, Belgium}


\abstract{%
 High-energy neutrinos are expected to be produced by the interaction of accelerated particles near the acceleration sites. For this reason, it is interesting to search for correlation in the arrival directions of ultra-high energy cosmic rays (UHECRs) and HE neutrinos. We present here the results of a search for correlations between UHECR events measured by the Pierre Auger Observatory and Telescope Array and high-energy neutrino candidate events from IceCube and ANTARES. We perform a cross-correlation analysis, where the angular separation between the arrival directions of UHECRs and neutrinos is scanned. When comparing the results with the expectations from a null hypothesis contemplating an isotropic distribution of neutrinos or of UHECR we obtain post-trial p-values of the order of $\sim 10^{-2}$.
}
\maketitle
\section{Introduction}
\label{intro}
The sources of ultra-high energy cosmic rays (UHECRs) are still unknown, mostly because of the difficulties in tracking their origin after they are deflected by magnetic fields. On the other hand, neutrinos, being neutral, do not suffer deflections and carry directional information on their origin. High-energy neutrinos are expected to be produced at the same sites of UHECRs, when the latter interact with the source itself or in its vicinity. It would be then natural to search for a correlation in the arrival directions of UHECRs and neutrinos. Previous searches performed by the Pierre Auger Observatory, Telescope Array and IceCube collaborations showed an interesting hint of correlation at a $3\sigma$ level \cite{jcap} which was later reduced with additional data \cite{ICRC2017}. In these proceedings we show the results of an update of this analysis, with new data from the original collaborations and with the addition of data from the ANTARES neutrino telescope. 
\section{The UHECR observatories and datasets}
\label{UHECRdata}
Ultra-high energy cosmic rays are observed via the extensive air showers (EAS) they induce in the atmosphere. The EAS can be observed in their longitudinal development through telescopes that detect the fluorescence emission of the atmosphere particles or sampled at the ground with arrays of particle detectors. The first measurement gives a direct calorimetric estimate of the energy of the primary particle but can be performed only during clear moonless nights (low duty cycle). The \textit{hybrid concept} that both the Pierre Auger Observatory and Telescope Array exploit relies on a cross-calibration of the surface detector with the events observed through fluorescence. In this way, an estimate of the energy of the primary particle can be performed with surface data only ($\sim 100 \%$ duty cycle) without relying on simulations. 
 A summary of UHECR data used in this and previous works can be found in Table \ref{tab_Data}.
\subsection{The Pierre Auger Observatory}
\label{Auger}
The Pierre Auger Observatory \cite{augernim,NIM2015}, located in Argentina at a latitude of $\sim -35^\circ$, is a $3000\;\text{km}^2$ cosmic-ray detector. It is a hybrid detector, using both a surface detector array (SD) composed of 1660 water-Cherenkov detectors and an atmospheric fluorescence detector (FD) made of
27 fluorescence telescopes to observe extensive air showers induced in the atmosphere by primary particles with energies $E>0.1\;\text{EeV}$\footnote{$1\;\text{EeV}=10^{18}\; \text{eV}$}.  For this study, 324 events with $E>52\;$EeV recorded by SD from January 2004 to April 2017 were used, resulting in 90 events more (corresponding to three years) with respect to previous analyses \cite{ICRC2017}. Events are observed up to a zenith angle of $80^\circ$, which translates into a field of view ranging from $-90^\circ$ to $+45^\circ$ in declination. The systematic uncertainty on the energy scale is 14\%, and the statistical uncertainty in the energy is smaller than 12\% for these events \cite{VerziICRC2013}, while the angular uncertainty is less than $0.9^\circ$ at these energies. 
\subsection{Telescope Array}
\label{TA}
Telescope Array (TA) \cite{TASDNIM,TAFDNIM} is a UHECR detector located in Utah, USA, at a latitude of $+39^\circ$. It is composed of 507 plastic scintillators distributed on a square grid covering a $700 \; \text{km}^2$ surface, overlooked by three fluorescence detector stations housing 38 telescopes. The events used for this analysis have an energy $E\geq 57 \:$EeV and zenith angles up to $55^\circ$. A total of 143 events recorded from May 2008 to May 2017 is used in this work, that corresponds to 34 more (two years) than the ones used in the latest update \cite{ICRC2017}.  These events have about $1.5^\circ$ angular resolution, $\sim20\%$ energy resolution is and a $\sim22\%$ systematic uncertainty on the energy scale\cite{TAspec}.
\section{The neutrino observatories and datasets}
\label{nudata}
High-energy astrophysical neutrinos are observed by instrumenting large volumes ($\sim 0.01-1\;\text{km}^3 $) of water or ice with photomultipliers (PMT) in order to use them as Cherenkov detectors. The shape of a neutrino event depends on the neutrino flavor and on the type of interaction it undergoes. Neutrinos of all flavors via neutral current (NC) and $\nu_e$ via charged current (CC) produce electromagnetic showers that are usually well-contained within the detector, offering a good energy estimate but a poor angular resolution (shower-like events, or \textit{cascades}). Vice-versa, muons produced in $\nu_\mu$ CC interactions tend to cross through the whole detector resulting in a good angular resolution and a poor energy estimate (track-like events, or \textit{tracks}). A summary of neutrino candidate events used in this and previous works can be found in Table \ref{tab_Data}.
\subsection{The ANTARES neutrino telescope}
\label{ANTARES}
The ANTARES neutrino telescope \cite{AntNIM} is located at $2475\;$m of depth in the Mediterranean sea, off the coast of Toulon, France (latitude: $+43^\circ$. It is composed of 12 $450\:$m-long strings, each equipped with 75 optical modules, at a depth of about $2500 \:$m.
The events used for this work are selected from the diffuse nine-year sample \cite{AntDiffuse} by requiring a ‘signalness’ $> 40\%$. The signalness is defined as the ratio of the number of expected astrophysical events over the sum of the expected atmospheric and astrophysical events for a given energy proxy. The selection resulted in two track-like events and no cascade events. The same signalness cut applied to the point-source sample \cite{AntPS} resulted in an additional track event selected, bringing the ANTARES dataset used in this analysis to a total of three tracks and no cascades. For these events, the median angular resolution is $\sim0.4^\circ$. 

\subsection{The IceCube South Pole Neutrino
Observatory}
\label{IC}
IceCube \cite{icecubeA,icecube} is a high-energy neutrino detector of $\sim1\;\text{km}^3$ composed by 86 strings instrumented by 5160 PMTs installed deep ($1.5-2.5\;$km) in the ice at the geographic South Pole (latitude: $-90^\circ$). The dataset used for this analysis is composed of events from two different samples. The main one is the six-year High-Energy Starting Events (HESE) dataset \cite{IC1ICRC2017}, which includes events generated by neutrinos of all flavors interacting inside the detection volume (starting events) with deposited energies ranging from 60 TeV up to 2 PeV. It includes 58 cascades and 15 tracks. The track sample is integrated with data from the dataset of through-going muons induced by charged current interactions of candidate $\nu_\mu$ from the Northern sky (diffuse through up-going tracks ) \cite{IceCubeMu}. This dataset adds 35 tracks corresponding to seven years of data from the eight-year data sample published in \cite{IC2ICRC2017}. Only events with ‘signalness’ $> 50\%$ are used (as described in Section \ref{ANTARES}), corresponding to a threshold of $\sim200 \:$TeV on the muon energy proxy.

Cascade events in the HESE sample have an angular resolution of $\sim 15^\circ $  above 100 TeV. Tracks are reconstructed with an angular resolution of $ \lesssim 1^\circ$. The resolution of the deposited energy for tracks and cascades is around
15\% \cite{IceCubeEnergy}, but cascades have a better resolution for the reconstructed neutrino energy since they deposit most of their energy inside the detector, while tracks do not.

\section{Analysis methods}
\label{method}
The search for correlations between UHECR and neutrino arrival directions is performed using different methods\cite{jcap}: a cross-correlation analysis, a likelihood analysis using neutrinos as stacked sources and a likelihood analysis using UHECRs as stacked sources to search for correlation with point-source neutrino samples (not described in paragraph \ref{nudata}). A recent update of the third analysis with an improved method is described in \cite{Schumacher2018}. In these proceedings an update of the cross-correlation analysis is presented.

The cross-correlation analysis is a simple yet completely model-independent analysis. It does not require assumptions on the deflection of UHECR in magnetic fields. This means that a part from the global energy cut chosen for the dataset by each collaboration, no energy information is needed on the single UHECR event. For this reason, no re-scaling of the UHECR energy scales to match the TA and Auger flux in the common field of view will be applied in the following, as it was done in previous works. Information on the flux matching of the two UHECR datasets can, however, be found in other joint works presented at this conference, in particular \cite{Ivanov:2018} and \cite{Biteau:2018}.

The principle of the cross-correlation analysis is to count the number $n_{OBS}$ of UHECR-neutrino pairs separated by less than a certain angular distance $\psi$. This number is then compared to the expected number of pairs $n_{EXP}$ separated by less than $\psi$ in the null-hypothesis scenario. For this analysis, two different null-hypotheses were investigated. The first one is an isotropic distribution of UHECR, obtained by generating isotropic datasets according to the UHECR exposure of the two experiments. The second null-hypothesis is the complementary one: an isotropic distribution of neutrinos, obtained by scrambling the real neutrino datasets, generating randomly right ascension values but keeping the original declination, thus preserving the declination-dependent acceptance of the neutrino observatories. The two null-hypotheses are investigated separately and thus when neutrino positions are randomized, UHECR ones are kept fixed and vice-versa. A large number of samples satisfying the null hypothesis are generated, and the fraction of generated samples where $n_{EXP}\geq n_{OBS}$ is observed is considered the local p-value. The analysis was performed at various $\psi$ values, ranging from $1^\circ$ to $30^\circ$ in $1^\circ$ steps in order to find the maximum departure from isotropy and the minimum local p-value. This p-value is then penalized taking into account the scan on $\psi$ and resulting in the final global p-value that will be reported in the following. The analysis was performed independently on the cascade and track neutrino datasets, thus obtaining four different p-values, one for each neutrino sample and each null hypothesis.

\section{Results}
\subsection{Previous results}
In \cite{jcap}, the first results of the analysis described in section \ref{method} were reported. No significant result was found using neutrino tracks, while with cascades the penalized p-values were $5 \times 10^{-4} $ with respect to an isotropic flux of UHECR and  $8.5  \times 10^{-3}$ with respect to an isotropic flux of neutrinos. In \cite{ICRC2017}, with additional neutrino data from IceCube and one year more of UHECR events from TA, the correlation with tracks continued to be not significant, while with cascades the p-values went up roughly one order of magnitude, respectively to $5.4 \times 10^{-3}$ and $1 \times 10^{-2}$.
\subsection{Updated results}
\begin{figure*}
\centering
\includegraphics[width=0.49\linewidth]{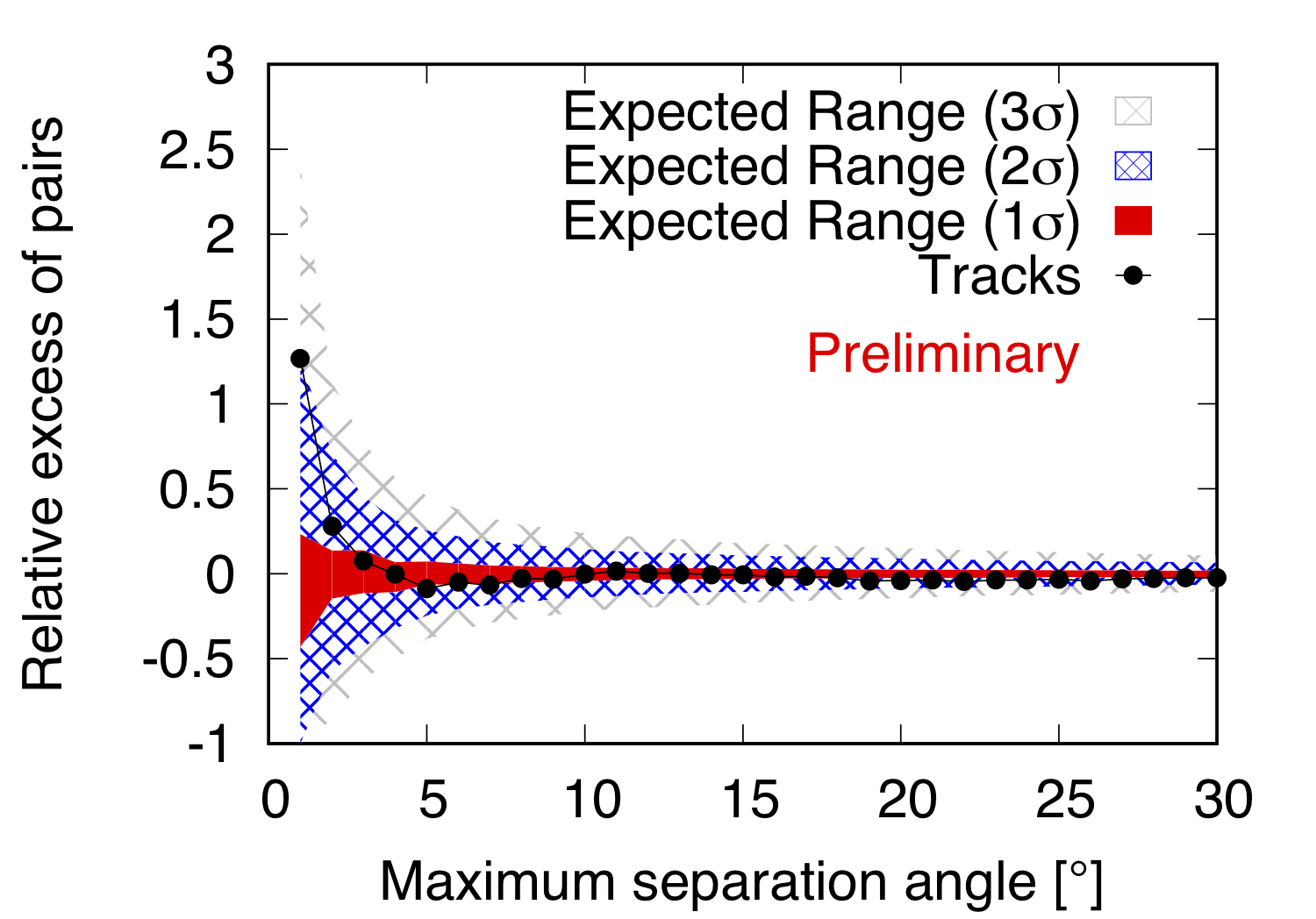} \includegraphics[width=0.49\linewidth]{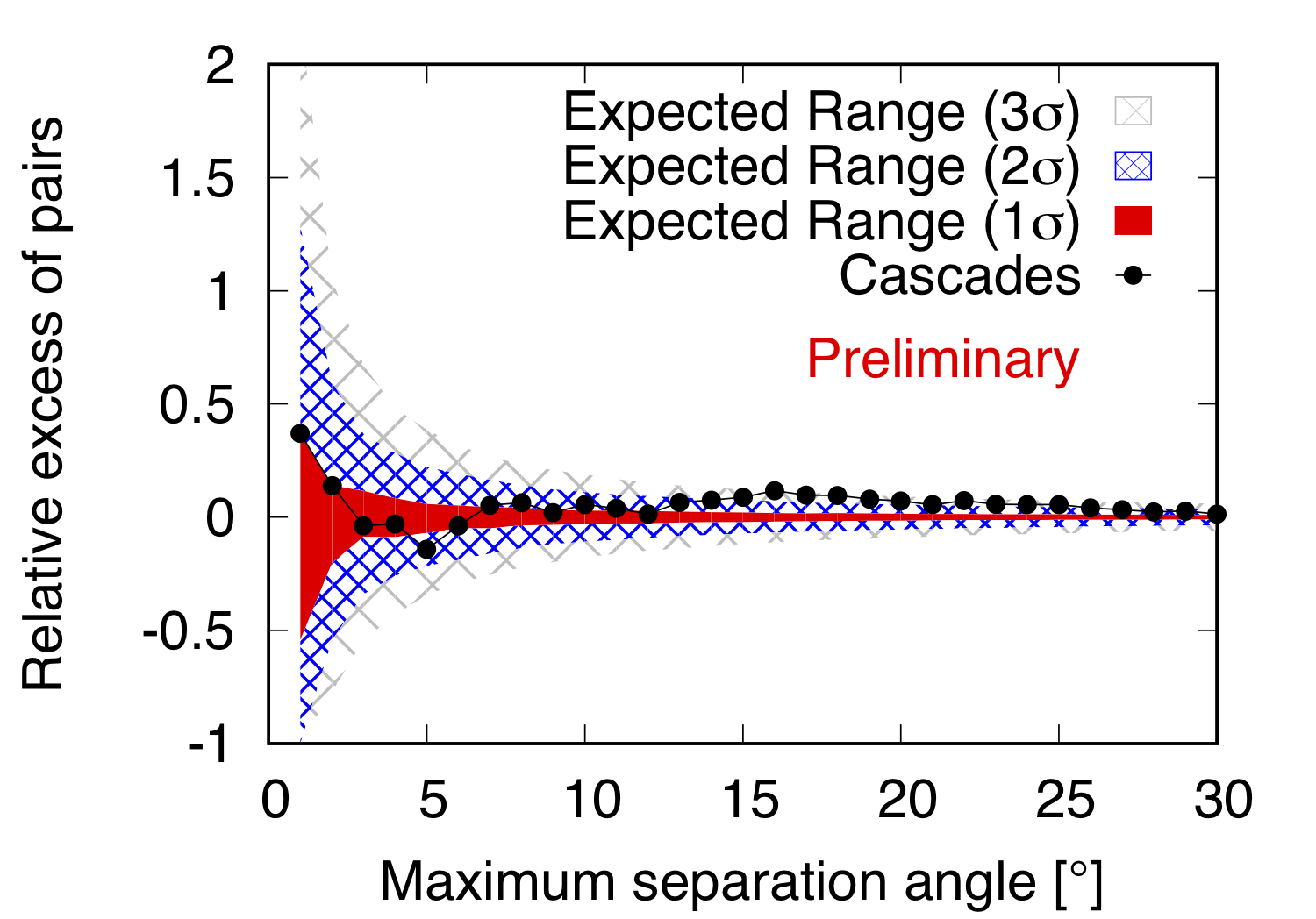}
\caption{Results of the cross-correlation analysis applied to the neutrino tracks (left) and cascades (right) samples. The relative excess of pairs is shown as a function of the maximum separation angle for data (dots) and expectation from an isotropic distribution of UHECR (colored area, with one, two and three-$\sigma$ dispersion). }
\label{res}       
\end{figure*}

We performed the analysis on the updated datasets as described in paragraph \ref{UHECRdata} and \ref{nudata}. The results of the scan in angle are shown in figure \ref{res}, where the relative excess of pairs with respect to the expected value from an isotropic distribution of UHECRs is shown for tracks (left) and cascades (right). In these conditions, the maximum departure from isotropy is found at $1^\circ$ for tracks, where four pairs are observed and at $16^\circ$ for tracks, with observed 623 pairs. The post-trial p-values are 0.45 for tracks and $2.7 \times 10^{-2}$ for cascades. The most significant angles are found to be the same for the null hypothesis of an isotropic distribution of neutrinos. The corresponding penalized p-values are 0.49 and $2.6 \times 10^{-2}$ for tracks and cascades respectively. A summary of the updated results compared to the previous ones can be found in table \ref{tab_Res}.

\section{Discussion}
In \cite{jcap}, a possible hint of a correlation between the arrival directions of UHECR and HE neutrinos was reported, using a simple and model-independent method. A first update published in \cite{ICRC2017} lowered the significance of the finding. This work uses new UHECR data from the Pierre Auger Observatory and Telescope Array experiment and new neutrino tracks from the ANTARES neutrino telescope. With these updated datasets, the post-trial p-values are now higher than 0.01 both considering as a null hypothesis an isotropic distribution of UHECRs or neutrinos.

This result can be explained by a number of considerations. First of all, the composition of ultra-high energy cosmic rays is not well known at these energies. This, together with the poor knowledge of the Galactic and extra-galactic magnetic field, implies that we cannot distinguish between UHECRs that suffer small or large deflections.  Secondly, as already pointed out in \cite{jcap}, the propagation of UHECRs is limited by the interactions with the cosmic backgrounds (GZK effects and similar): depending on the composition, it is expected that cosmic rays at these energies cannot arrive from further than 10-100 Mpc. On the other hand, neutrinos can reach us from cosmological distances so that in principle only a few percent of neutrinos would be expected from the sources within this horizon and thus in correlation with the arrival directions of the detected UHECRs. Finally, neutrinos originated at cosmic rays acceleration sites are expected to carry few percents of the energy of the original cosmic-ray. Thus, the neutrinos observed by IceCube and ANTARES would have been produced by cosmic rays of much lower energy than the ones in the datasets by Auger and TA. For this reason, it is possible that only a fraction of the neutrinos in our data could come from a site that is actually capable of accelerating cosmic rays up to $\sim10^{20}\;$eV. We expect to carry on this work by soon updating the other two analyses with the latest data.


\begin{table*}
\centering
\caption{Summary of the data used in this work compared to previous ones.}
\label{tab_Data}       
\begin{tabular}{l||c|c|c}
                     & JCAP 2016 \cite{jcap}                                                                                         & ICRC 2017   \cite{ICRC2017}                                                               & \textbf{\begin{tabular}[c]{@{}c@{}}UHECR 2018 \\ (This work)\end{tabular}}                           \\ \hline \hline
Auger UHECR          & \multicolumn{2}{c|}{\begin{tabular}[c]{@{}c@{}}231 events\\ \footnotesize{1/1/2004 - 31/3/2014}\end{tabular}}                                                                                 & \textbf{\begin{tabular}[c]{@{}c@{}}324 events\\ \footnotesize{1/1/2004 - 30/4/2017}\end{tabular}}                   \\ \hline
TA UHECR             & \begin{tabular}[c]{@{}c@{}}87 events\\ \footnotesize{11/5/2008 - 1/5/2014}\end{tabular}                           & \begin{tabular}[c]{@{}c@{}}109 events\\ \footnotesize{11/5/2008 - 1/5/2015}\end{tabular} & \textbf{\begin{tabular}[c]{@{}c@{}}143 events\\ \footnotesize{11/5/2008 - 1/5/2017}\end{tabular}}                   \\ \hline \hline
IceCube HE $\nu$ & \begin{tabular}[c]{@{}c@{}}39 cascades\\ 16 tracks \\ \footnotesize{(4 yrs HESE + diffuse up-going)}\end{tabular} & \multicolumn{2}{c}{\textbf{\begin{tabular}[c]{@{}c@{}}58 cascades\\ 49 tracks\\ \footnotesize{(6 yrs HESE + diffuse up-going)} \end{tabular}}}                                                                                    \\ \hline
ANTARES HE $\nu$ & \multicolumn{2}{c|}{-}                                                                                                                                                         & \textbf{\begin{tabular}[c]{@{}c@{}}3 tracks\\ \footnotesize{(9-years diffuse + point source )}\end{tabular}}
\end{tabular}
\end{table*}

\begin{table*}
\centering
\caption{Summary of the post-trial p-values obtained in this work compared to previous ones.}
\label{tab_Res}       
\begin{tabular}{l||c|c|c}
                     & JCAP 2016 \cite{jcap}                                                                                         & ICRC 2017 \cite{ICRC2017}                                                                & \textbf{\begin{tabular}[c]{@{}c@{}}UHECR 2018 \\ (This work)\end{tabular}}                           \\ \hline \hline
tracks wrt an isotropic 
flux of UHECR           &  0.28 & 0.48 &\textbf{ 0.45} \\ \hline
tracks wrt an isotropic 
flux of neutrinos             &   &0.52&\textbf{0.49} \\ \hline \hline
cascades wrt an isotropic flux of UHECR  & $ 5 \times 10^{-4}$& $5.4 \times 10^{-3}$ & $\boldsymbol{2.7 \times 10^{-2}}$
  \\ \hline
cascades wrt an isotropic flux of neutrinos $\nu$ & $8.5 \times 10^{-3}$ &$1.0 \times 10^{-2} $& $\boldsymbol{2.6 \times 10^{-2}}$

\end{tabular}
\end{table*}
%
\bibliography{UHECRbib.bib}
%
%
%
%





\end{document}